\documentclass[sigconf,natbib,9pt]{acmart}

\usepackage{booktabs} 
\usepackage{graphicx}
\usepackage{url}
\usepackage{microtype}
\usepackage{tabularx,amsmath}
\usepackage{amssymb}
\usepackage{multirow}
\usepackage{color}
\usepackage[normalem]{ulem}
\usepackage[english]{babel}
\usepackage{colortbl}
\usepackage[noend]{algorithmic}
\usepackage{algorithm}
\usepackage{paralist}
\usepackage{enumerate}
\usepackage{enumitem}
\usepackage{acronym}
\usepackage[skip=0pt]{caption}
\usepackage{accents}
\usepackage{setspace}
\usepackage{balance}

\setcitestyle{square,comma,numbers,sort&compress,sectionbib}

\DeclareMathOperator{\sgn}{sgn}

\newbox\itembox
\def\itemlistlabel#1{#1\hfill}
\def\itemlist#1{\setbox\itembox=\hbox{#1}%
                \list{}{\labelwidth\wd\itembox
                             \leftmargin\labelwidth
                             \advance\leftmargin by\itemindent
                             \advance\leftmargin by\labelsep
                             \let\makelabel\itemlistlabel}}

\newcommand{\enkelop}{$^{\vartriangle}$}
\newcommand{\dubbelop}{$^{\blacktriangle}$}
\newcommand{\enkelneer}{$^{\triangledown}$}
\newcommand{\dubbelneer}{$^{\blacktriangledown}$}

\newcommand{\ubar}[1]{\underaccent{\bar}{#1}}

\acrodef{IR}{Information Retrieval}
\acrodef{LTR}{Learning to Rank}
\acrodef{OLTR}{Online Learning to Rank}
\acrodef{TDM}{Team-Draft Multileaving}
\acrodef{OM}{Optimized Multileaving}
\acrodef{PM}{Probabilistic Multileaving}
\acrodef{SOSM}{Sample-Scored-Only Multileaving}
\acrodef{PPM}{Pairwise Preference Multileaving}
\acrodef{LM}{Lambda Multileaving}
\acrodef{PBI}{Preference-Based Balanced Interleaving}
\acrodef{TDI}{Team Draft Interleaving}
\acrodef{OI}{Optimized Interleaving}
\acrodef{PI}{Probabilistic Interleaving}

\newif\ifanon
\ifx\anonymous\undefined
  \anonfalse
\else
  \anontrue
\fi

\title[Online Evaluation with Theoretical Guarantees]{Sensitive and Scalable Online Evaluation\\ with Theoretical Guarantees}

\author{Harrie Oosterhuis}
\orcid{}
\affiliation{%
\institution{University of Amsterdam}
\city{Amsterdam}
\country{The Netherlands}
}
\email{oosterhuis@uva.nl}

\author{Maarten de Rijke}
\orcid{0000-0002-1086-0202}
\affiliation{%
\institution{University of Amsterdam}
\city{Amsterdam}
\country{The Netherlands}
}
\email{derijke@uva.nl}

\begin{document}

\begin{abstract}
Multileaved comparison methods generalize interleaved comparison methods to provide a scalable approach for comparing ranking systems based on regular user interactions. Such methods enable the increasingly rapid research and development of search engines.
However, existing multileaved comparison methods that provide reliable outcomes do so by degrading the user experience during evaluation. Conversely, current multileaved comparison methods that maintain the user experience cannot guarantee correctness.
Our contribution is two-fold.
First, we propose a theoretical framework for systematically comparing multileaved comparison methods using the notions of \emph{considerateness}, which concerns maintaining the user experience, and \emph{fidelity}, which concerns reliable correct outcomes.
Second, we introduce a novel multileaved comparison method, \ac{PPM}, that performs comparisons based on document-pair preferences, and prove that it is \emph{considerate} and has \emph{fidelity}. We show empirically that, compared to previous multileaved comparison methods, \acs{PPM} is more \emph{sensitive} to user preferences and \emph{scalable} with the number of rankers being compared.
\end{abstract}

%

\copyrightyear{2017} 
\acmYear{2017} 
\setcopyright{acmlicensed}
\acmConference{CIKM'17}{}{November 6--10, 2017, Singapore.}
\acmPrice{15.00}
\acmDOI{https://doi.org/10.1145/3132847.3132895}
\acmISBN{ISBN 978-1-4503-4918-5/17/11}

\maketitle


\section{Introduction}
\label{sec:intro}

Evaluation is of tremendous importance to the development of modern search engines. Any proposed change to the system should be verified to ensure it is a true improvement.
Online approaches to evaluation aim to measure the actual utility of an \ac{IR} system in a natural usage
environment~\citep{hofmann-online-2016}. 
Interleaved comparison methods are a within-subject setup for online experimentation in \ac{IR}. For interleaved comparison, two experimental conditions (``control'' and ``treatment'') are typical. Recently, multileaved comparisons have been introduced for the purpose of efficiently comparing large numbers of rankers~\citep{Schuth2014a,brost2016improved}. These multileaved comparison methods were introduced as an extension to interleaving and the majority are directly derived from their interleaving counterparts \cite{Schuth2014a, schuth2015probabilistic}. The effectiveness of these methods has thus far only been measured using simulated experiments on public datasets. While this gives some insight into the general \emph{sensitivity} of a method, there is no work that assesses under what circumstances these methods provide correct outcomes and when they break. Without knowledge of the theoretical properties of multileaved comparison methods we are unable to identify when their outcomes are reliable.

In prior work on interleaved comparison methods a theoretical framework has been introduced that provides explicit requirements that an interleaved comparison method should satisfy~\citep{hofmann2013fidelity}. We take this approach as our starting point and adapt and extend it to the setting of multileaved comparison methods. Specifically, the notion of \emph{fidelity} is central to \citet{hofmann2013fidelity}'s previous work; Section~\ref{sec:multileaving} describes the framework with its requirements of \emph{fidelity}. In the setting of multileaved comparison methods, this means that a multileaved comparison method should always recognize an unambiguous winner of a comparison. We also introduce a second notion, \emph{considerateness}, which says that a comparison method should not degrade the user experience, e.g., by allowing all possible permutations of documents to be shown to the user. In this paper we examine all existing multileaved comparison methods and find that none satisfy both the \emph{considerateness} and \emph{fidelity} requirements. In other words, no existing multileaved comparison method is correct without sacrificing the user experience. 

To address this gap, we propose a novel multileaved comparison method, \acf{PPM}. \ac{PPM} differs from existing multileaved comparison methods as its comparisons are based on inferred pairwise document preferences, whereas existing multileaved comparison methods either use some form of document assignment \cite{Schuth2014a, schuth2015probabilistic} or click credit functions \cite{Schuth2014a, brost2016improved}. We prove that \ac{PPM} meets both the \emph{considerateness} and the \emph{fidelity} requirements, thus \ac{PPM} guarantees correct winners in unambiguous cases while maintaining the user experience at all times. Furthermore, we show empirically that \ac{PPM} is more \emph{sensitive} than existing methods, i.e., it makes fewer errors in the preferences it finds. Finally, unlike other multileaved comparison methods, \ac{PPM} is computationally efficient and \emph{scalable}, meaning that it maintains most of its \emph{sensitivity} as the number of rankers in a comparison increases.

In this paper we address the following research questions:
 \begin{enumerate}[label={\bf RQ\arabic*},leftmargin=*,nosep]
    \item Does \ac{PPM} meet the \emph{fidelity} and \emph{considerateness} requirements?\label{rq:theory}
    \item Is \ac{PPM} more sensitive than existing methods when comparing multiple rankers?\label{rq:sensitive}
\end{enumerate}
To summarize, our contributions are:
 \begin{enumerate}[nosep,leftmargin=14pt]
    \item A theoretical framework for comparing multileaved comparison methods;
    \item A comparison of all existing multileaved comparison methods in terms of \emph{considerateness}, \emph{fidelity} and \emph{sensitivity};
    \item A novel multileaved comparison method that is \emph{considerate} and has \emph{fidelity} and is more \emph{sensitive} than existing methods.
\end{enumerate}

\section{Related Work}
\label{sec:related}

Evaluation of information retrieval systems is a core problem in IR. Two types of approach are common to designing reliable methods for measuring an IR system's effectiveness.  Offline approaches such as the Cranfield paradigm~\cite{sanderson-test-2010} are effective for measuring topical relevance, but have difficulty taking into account contextual information including the user's current situation, fast
changing information needs, and past interaction history with the
system~\citep{hofmann-online-2016}. In contrast, online approaches to evaluation aim to measure
the actual utility of an \ac{IR} system in a natural usage
environment. User feedback in online
evaluation is usually implicit, in the form of clicks, dwell time, etc.

By far the most common type of controlled experiment on the web is A/B testing~\citep{kohavi2009controlled,Kohavi2013}. This is a classic between-subject experiment, where each subject is exposed to one of two conditions, \emph{control}---the current system---and \emph{treatment}---an experimental system that is assumed to outperform the control.

An alternative experiment design uses a within-subject setup, where all study participants are exposed to both experimental conditions. Interleaved comparisons \cite{Joachims2002,radlinski08:how} have been developed specifically for online experimentation in IR. Interleaved comparison methods have two main ingredients. First, a method for constructing interleaved result lists specifies how to select documents from the original rankings (``control'' and ``treatment''). Second, a method for inferring comparison outcomes based on observed user interactions with the interleaved result list. Because of their within-subject nature, interleaved comparisons can be up to two orders of magnitude more efficient than A/B tests in effective sample size for studies
of comparable dependent variables~\citep{chapelle2012large}.

For interleaved comparisons, two experimental conditions are typical. Extensions to multiple conditions have been introduced by \citet{Schuth2014a}. Such \emph{multileaved} comparisons are an efficient online evaluation method for comparing multiple rankers simultaneously. Similar to interleaved comparison methods~\cite{hofmann2011probabilistic, radlinski2013optimized, radlinski08:how, joachims03:evaluating}, a multileaved comparison infers preferences between rankers. Interleaved comparisons do this by presenting users with interleaved result lists; these represent two rankers in such a way that a preference between the two can be inferred from clicks on their documents. Similarly, for multileaved comparisons multileaved result lists are created that allow more than two rankers to be represented in the result list. As a consequence, multileaved comparisons can infer preferences between multiple rankers from a single click. Due to this property multileaved comparisons require far fewer interactions than interleaved comparisons to achieve the same accuracy when multiple rankers are involved \cite{Schuth2014a,  schuth2015probabilistic}.

\begin{algorithm}[h]
\begin{algorithmic}[1]
\STATE \textbf{Input}: set of rankers $\mathcal{R}$, documents $D$, no. of timesteps $T$.
\STATE $P \leftarrow \mathbf{0}$ \hfill\textit{\small// initialize $|\mathcal{R}|\times |\mathcal{R}|$ preference matrix} \label{alg:genmul:init}
\FOR{$t = 1,\ldots ,T$}
    \STATE $q_t \leftarrow \textit{wait\_for\_user()}$ \hfill \textit{\small // receive query from user} \label{alg:genmul:query}
    \FOR{$i = 1,\ldots,|\mathcal{R}|$}
    	\STATE $\mathbf{l}_i \leftarrow \mathbf{r}_i(q,D)$ \hfill \textit{\small // create ranking for query per ranker}\label{alg:genmul:rank}
    \ENDFOR
    \STATE $\mathbf{m}_t \leftarrow \textit{combine\_lists}(\mathbf{l}_1,\ldots,\mathbf{l}_{R})$\hfill\textit{\small// combine into multileaved list} \label{alg:genmul:multileaving}
    \STATE $\mathbf{c} \leftarrow \textit{display}(\mathbf{m}_t)$ \hfill \textit{\small // display to user and record interactions} \label{alg:genmul:display}
    \FOR{$i = 1,\ldots,|\mathcal{R}|$}
    	\FOR{$j = 1,\ldots,|\mathcal{R}|$}
		\STATE $P_{ij} \leftarrow P_{ij} + \textit{infer}( i,j, \mathbf{c}, \mathbf{m}_t)$ \hfill \textit{\small // infer pref. between rankers}
		\label{alg:genmul:infer}
	\ENDFOR
    \ENDFOR
\ENDFOR
\RETURN $P$
\end{algorithmic}
\caption{General pipeline for multileaved comparisons.}
\label{alg:generalmultileaving}
\end{algorithm}

The general approach for every multileaved comparison method is described in Algorithm~\ref{alg:generalmultileaving}; here, a comparison of a set of rankers $\mathcal{R}$ is performed over $T$ user interactions. After the user submits a query $q$ to the system (Line~\ref{alg:genmul:query}), a ranking $\mathbf{l}_i$ is generated for each ranker $\mathbf{r}_i$ in $\mathcal{R}$ (Line~\ref{alg:genmul:rank}). These rankings are then combined into a single result list by the multileaving method (Line~\ref{alg:genmul:multileaving}); we refer to the resulting list $\mathbf{m}$ as the  multileaved result list. In theory a multileaved result list could contain the entire document set, however in practice a length $k$ is chosen beforehand, since users generally only view a restricted number of result pages. This multileaved result list is presented to the user who has the choice to interact with it or not. Any interactions are recorded in $\mathbf{c}$ and returned to the system (Line~\ref{alg:genmul:display}). While $\mathbf{c}$ could contain any interaction information~\cite{kharitonov2015generalized}, in practice multileaved comparison methods only consider clicks. Preferences between the rankers in $\mathcal{R}$ can be inferred from the interactions and the preference matrix $P$ is updated accordingly (Line~\ref{alg:genmul:infer}). The method of inference (Line~\ref{alg:genmul:infer}) is defined by the multileaved comparison method (Line~\ref{alg:genmul:multileaving}). By aggregating the inferred preferences of many interactions a multileaved comparison method can detect preferences of users between the rankers in $\mathcal{R}$. Thus it provides a method of evaluation without requiring a form of explicit annotation.

By instantiating the general pipeline for multileaved comparisons shown in Algorithm~\ref{alg:generalmultileaving}, i.e., the combination method at Line~\ref{alg:genmul:rank} and the inference method at Line~\ref{alg:genmul:infer}, we obtain a specific multileaved comparison method. We detail all known multileaved comparison methods in Section~\ref{sec:existingmethods} below. 

\smallskip\noindent%
What we add on top of the work discussed above is a theoretical framework that allows us to assess and compare multileaved comparison methods. In addition, we propose an accurate and scalable multileaved comparison method that is the only one to satisfy the properties specified in our theoretical framework and that also proves to be the most efficient multileaved comparison method in terms of much reduced data requirements.


\section{A Framework for Assessing Multileaved Comparison Methods}
\label{sec:multileaving}

Before we introduce a novel multileaved comparison method in Section~\ref{sec:novelmethod}, we propose two theoretical requirements for multileaved comparison methods. These theoretical requirements will allow us to assess and compare existing multileaved comparison methods. Specifically, we introduce two theoretical properties: \emph{considerateness} and \emph{fidelity}. These properties guarantee correct outcomes in \emph{unambigious} cases while always maintaining the user experience. In Section~\ref{sec:existingmethods} we show that no currently available multileaved comparison method satisfies both properties. This motivates the introduction of a method that satisfies both properties in Section~\ref{sec:novelmethod}. 

\subsection{Considerateness}
Firstly, one of the most important properties of a multileaved comparison method is how \textbf{considerate} it is. Since evaluation is done online it is important that the search experience is not substantially altered \citep{Joachims2002, radlinski2013optimized}. In other words, users should not be obstructed to perform their search tasks during evaluation. As maintaining a user base is at the core of any search engine, methods that potentially degrade the user experience are generally avoided. Therefore, we set the following requirement: the displayed multileaved result list should never show a document $d$ at a rank $i$ if every ranker in $\mathcal{R}$ places it at a lower rank. Writing $r(d,\mathbf{l}_j)$ for the rank of $d$ in the ranking $\mathbf{l}_j$ produced by ranker $\mathbf{r}_j$, this boils down to:
\begin{align}
\mathbf{m}_i = d \rightarrow \exists \mathbf{r}_j \in \mathcal{R},  r(d,\mathbf{l}_j) \leq i. \label{eq:obstruction}
\end{align}
Requirement~\ref{eq:obstruction} guarantees that a document can never be displayed higher than any ranker would. In addition, it guarantees that if all rankers agree on the top $n$ documents, the resulting multileaved result list $\mathbf{m}$ will display the same top $n$.

\subsection{Fidelity}
Secondly, the preferences inferred by a multileaved comparison method should correspond with those of the user with respect to retrieval quality, and should be robust to user behavior that is unrelated to retrieval quality \cite{Joachims2002}. In other words, the preferences found should be correct in terms of ranker quality. However, in many cases the relative quality of rankers is unclear. For that reason we will use the notion of \textbf{fidelity} \cite{hofmann2013fidelity} to compare the correctness of a multileaved comparison method. \emph{Fidelity} was introduced by \citet{hofmann2013fidelity} and describes two general cases in which the preference between two rankers is unambiguous. To have \emph{fidelity} the expected outcome of a method is required to be correct in all matching cases. However, the original notion of \emph{fidelity} only considers two rankers as it was introduced for interleaved comparison methods, therefore the definition of \emph{fidelity} must be expanded to the multileaved case. First we describe the following concepts:

\paragraph{Uncorrelated clicks} Clicks are considered \emph{uncorrelated} if relevance has no influence on the likelihood that a document is clicked.
We write $r(d_i,\mathbf{m})$ for the rank of document $d_i$ in multileaved result list $\mathbf{m}$ and $P(\mathbf{c}_l\mid q, \mathbf{m}_l=d_i)$ for the probability of a click at the rank $l$ at which $d_i$ is displayed: $l = r(d_i,\mathbf{m})$. Then, for a given query $q$
\begin{equation}
\begin{aligned}
& \textit{uncorrelated}(q) \Leftrightarrow {}\\
&\qquad \forall l, \forall d_{i,j}, P(\mathbf{c}_l\mid q, \mathbf{m}_l=d_i) = P(\mathbf{c}_l\mid q, \mathbf{m}_l=d_j).
\end{aligned}
\label{eq:uncorrelated} 
\end{equation}

\paragraph{Correlated clicks} We consider clicks correlated if there is a positive correlation between document relevance and clicks. However we differ from \citet{hofmann2013fidelity} by introducing a variable $k$ that denotes at which rank users stop considering documents. Writing $P(\mathbf{c}_i\mid \textit{rel}(\mathbf{m}_i,q))$ for the probability of a click at rank $i$ if a document relevant to query $q$ is displayed at this rank, we set
\begin{equation}
\begin{aligned}
& \textit{correlated}(q, k) \Leftrightarrow {}  \forall i \geq k, P(\mathbf{c}_i) = 0  {}\\
& \qquad \land \forall i < k, P(\mathbf{c}_i\mid \textit{rel}(\mathbf{m}_i,q)) > P(\mathbf{c}_i \mid \neg \textit{rel}(\mathbf{m}_i,q)).
\end{aligned}
\label{eq:correlated}
\end{equation}
Thus under correlated clicks a relevant document is more likely to be clicked than a non-relevant one at the same rank, if they appear above rank $k$.

\paragraph{Pareto domination} Ranker $\mathbf{r}_1$ \emph{Pareto dominates} ranker $\mathbf{r}_2$ if all relevant documents are ranked at least as high by $\mathbf{r}_1$ as by $\mathbf{r}_2$ and $\mathbf{r}_1$ ranks at least one relevant document higher. Writing $\mathit{rel}$ for the set of relevant documents that are ranked above $k$ by at least one ranker, i.e., $\mathit{rel} = \{d \mid \mathit{rel}(d,q) \land \exists \mathbf{r}_n \in \mathcal{R}, r(d,\mathbf{l}_n) > k\}$,
we require that the following holds for every query $q$ and any rank $k$:
\begin{equation}
\begin{aligned}
&\textit{Pareto}( \mathbf{r}_i, \mathbf{r}_j, q, k) \Leftrightarrow {}\\
&\qquad\forall d \in \textit{rel}, r(d, \mathbf{l}_i) \leq r(d, \mathbf{l}_j) \land \exists d \in \textit{rel}, r(d, \mathbf{l}_i) < r(d, \mathbf{l}_j).
\end{aligned}
\end{equation}

\noindent%
Then, \emph{fidelity} for multileaved comparison methods is defined by the following two requirements:
\begin{enumerate}[nosep,leftmargin=14pt]
\item \label{fidelity:unbias}
Under uncorrelated clicks the expected outcome may find no preferences between any two rankers in $\mathcal{R}$:
\begin{align}
\forall q, \forall (\mathbf{r}_i, \mathbf{r}_j) \in \mathcal{R}, \mathit{uncorrelated}(q) \Rightarrow E[P_{ij}\mid q] = 0.
\end{align}
\item  \label{fidelity:pareto}
Under correlated clicks, a ranker that Pareto dominates all other rankers must win the multileaved comparison in expectation:
\begin{equation}
\begin{aligned}
&\forall k, \forall q, \forall \mathbf{r}_{i} \in \mathcal{R}, \big(\mathit{correlated}(q,k) \land \\
& \qquad\qquad \forall \mathbf{r}_{j} \in \mathcal{R}, i \not = j \rightarrow \mathit{Pareto}( \mathbf{r}_i, \mathbf{r}_j, q,k)\big)\\
&\qquad \Rightarrow  \left(\forall \mathbf{r}_{j} \in \mathcal{R}, i \not = j \rightarrow  E[P_{ij}\mid q] > 0\right).
\end{aligned}
\end{equation}
\end{enumerate}
Note that for the case where $|\mathcal{R}| = 2$ and if only $k = |D|$ is considered, these requirements are the same as for interleaved comparison methods~\citep{hofmann2013fidelity}.
The $k$ parameter was added to allow for \emph{fidelity} in \emph{considerate} methods, since it is impossible to detect preferences at ranks that users never consider without breaking the \emph{considerateness} requirement. We argue that differences at ranks that users are not expected to observe should not affect comparison outcomes.
 \emph{Fidelity} is important for a multileaved comparison method as it ensures that an unambiguous winner is expected to be identified. Additionally, the first requirement ensures unbiasedness when clicks are unaffected by relevancy.

\subsection{Additional properties}
\label{sec:non-theory}
In addition to the two theoretical properties listed above, considerateness and fidelity, we also scrutinize multileaved comparison methods to determine whether they accurately find preferences between all rankers in $\mathcal{R}$ and minimize the number of user impressions required do so. This empirical property is commonly known as \textbf{sensitivity}~\citep{Schuth2014a, hofmann2013fidelity}. In Section~\ref{sec:experiments} we describe experiments that are aimed at comparing the sensitivity of multileaved comparison methods. Here, two aspects of every comparison are considered: the level of error at which a method converges and the number of impressions required to reach that level. Thus, an interleaved comparison method that learns faster initially but does not reach the same final level of error is deemed worse. 


\section{An Assessment of Existing Multileaved Comparison Methods}
\label{sec:existingmethods}

We briefly examine all existing multileaved comparison methods to determine whether they meet the \emph{considerateness} and \emph{fidelity} requirements. An investigation of the empirical sensitivity requirement is postponed until Section~\ref{sec:experiments}~and~\ref{sec:results}. 

\subsection{Team Draft Multileaving}


\ac{TDM} was introduced by \citet{Schuth2014a} and is based on the previously proposed \ac{TDI} \cite{radlinski08:how}. Both methods are inspired by how team assignments are often chosen for friendly sport matches.
The multileaved result list is created by sequentially sampling rankers without replacement; the first sampled ranker places their top document at the first position of the multileaved list.
Subsequently, the next sampled ranker adds their top pick of the remaining documents. When all rankers have been sampled, the process is continued by sampling from the entire set of rankers again. The method is stops when all documents have been added. When a document is clicked, \ac{TDM} assigns the click to the ranker that contributed the document. For each impression binary preferences are inferred by comparing the number of clicks each ranker received.

It is clear that \ac{TDM} is \emph{considerate} since each added document is the top pick of at least one ranker. However, \ac{TDM} does not meet the fidelity requirements. This is unsurprising as previous work has proven that \ac{TDI} does not meet these requirements \cite{radlinski2013optimized,hofmann2011probabilistic,hofmann2013fidelity}. Since \ac{TDI} is identical to \ac{TDM} when the number of rankers is $|\mathcal{R}| = 2$, \ac{TDM} does not have \emph{fidelity} either.

\subsection{Optimized Multileaving}


\ac{OM} was proposed by \citet{Schuth2014a} and serves as an extension of \ac{OI} introduced by \citet{radlinski2013optimized}. The allowed multileaved result lists of \ac{OM} are created by sampling rankers with replacement at each iteration and adding the top document of the sampled ranker. However, the probability that a multileaved result list is shown is not determined by the generative process. Instead, for a chosen credit function \ac{OM} performs an optimization that computes a probability for each multileaved result list so that the expected outcome is unbiased and sensitive to correct preferences.

All of the allowed multileaved result lists of \ac{OM} meet the \emph{considerateness} requirement, and in theory instantiations of \ac{OM} could have \emph{fidelity}. However, in practice \ac{OM} does not meet the \emph{fidelity} requirements. There are two main reasons for this. First, it is not guaranteed that a solution exists for the optimization that \ac{OM} performs. For the interleaving case this was proven empirically when $k=10$ \cite{radlinski2013optimized}. However, this approach does not scale to any number of rankers. Secondly, unlike \ac{OI}, \ac{OM} allows more result lists than can be computed in a feasible amount of time. Consider the top $k$ of all possible multileaved result lists; in the worst case this produces $|\mathcal{R}|^k$ lists. Computing all lists for a large value of $|\mathcal{R}|$ and performing linear constraint optimization over them is simply not feasible. As a solution, \citet{Schuth2014a} propose a method that samples from the allowed multileaved result lists and relaxes constraints when there is no exact solution. Consequently, there is no guarantee that this method does not introduce bias. Together, these two reasons show that the \emph{fidelity} of \ac{OI} does not imply fidelity of \ac{OM}. It also shows that \ac{OM} is  computationally very costly. 

\subsection{Probabilistic Multileaving}
\ac{PM} \citep{schuth2015probabilistic} is an extension of \ac{PI} \cite{hofmann2011probabilistic}, which was designed to solve the flaws of \ac{TDI}. Unlike the previous methods, \ac{PM} considers every ranker as a distribution over documents, which is created by applying a soft-max to each of them. A multileaved result list is created by sampling a ranker with replacement at each iteration and sampling a document from the ranker that was selected. After the sampled document has been added, all rankers are renormalized to account for the removed document. During inference \ac{PM} credits every ranker the expected number of clicked documents that were assigned to them. This is done by marginalizing over the possible ways the list could have been constructed by \ac{PM}. A benefit of this approach is that it allows for comparisons on historical data \cite{hofmann2011probabilistic, hofmann2013fidelity}.

A big disadvantage of \ac{PM} is that it allows any possible ranking to be shown, albeit not with uniform probabilities. This is a big deterrent for the usage of \ac{PM} in operational settings. Furthermore, it also means that \ac{PM} does not meet the \emph{considerateness} requirement. On the other hand, \ac{PM} does meet the \emph{fidelity} requirements, the proof for this follows from the fact that every ranker is equally likely to add a document at each location in the ranking. Moreover, if multiple rankers want to place the same document somewhere they have to share the resulting credits.\footnote{\citet{brost2016improved} proved that if the preferences at each impression are made binary the  \emph{fidelity} of \ac{PM} is lost.} Similar to \ac{OM}, \ac{PM} becomes infeasible to compute for a large number of rankers $|\mathcal{R}|$; the number of assignments in the worst case is $|R|^{k}$. Fortunately, \ac{PM} inference can be estimated by sampling assignments in a way that maintains \emph{fidelity} \cite{schuth2015probabilistic, oosterhuis2016probabilistic}.

\subsection{Sample Only Scored Multileaving}
\ac{SOSM} was introduced by \citet{brost2016improved} in an attempt to create a more scalable multileaved comparison method. It is the only existing multileaved comparison method that does not have an interleaved comparison counterpart.
\ac{SOSM} attempts to increase \emph{sensitivity} by ignoring all non-sampled documents during inference. Thus, at each impression a ranker receives credits according to how it ranks the documents that were sampled for the displayed multileaved result list of size $k$. The preferences at each impression are made binary before being added to the mean. \ac{SOSM} creates multileaved result lists following the same procedure as \ac{TDM}, a choice that seems arbitrary.

\ac{SOSM} meets the \emph{considerateness} requirements for the same reason \ac{TDM} does. However, \ac{SOSM} does not meet the fidelity requirement. We can prove this by providing an example where preferences are found under uncorrelated clicks. Consider the two documents \textit{A} and \textit{B} and the three rankers with the following three rankings:
\begin{align*}
\mathbf{l}_1 = \textit{AB}, \qquad
\mathbf{l}_2 = \mathbf{l}_3 = \textit{BA}. 
\end{align*}
The first requirement of fidelity states that under uncorrelated clicks no preferences may be found in expectation. Uncorrelated clicks are unconditioned on document relevance (Equation~\ref{eq:uncorrelated}); however, it is possible that they display position bias \cite{yue2010beyond}. Thus the probability of a click at the first rank may be greater than at the second:
\begin{align*}
P(\mathbf{c}_1\mid q) > P(\mathbf{c}_2\mid q).
\end{align*}
Under position biased clicks the expected outcome for each possible multileaved result list is not zero. For instance, the following preferences are expected:
\begin{align}
E[P_{12}\mid \mathbf{m} = \textit{AB}] &> 0, \nonumber\\
E[P_{12}\mid \mathbf{m} = \textit{BA}] &< 0, \nonumber\\
E[P_{12}\mid \mathbf{m} = \textit{AB}] &= -E[P_{12}\mid \mathbf{m} = \textit{BA}]. \nonumber
\end{align}
Since \ac{SOSM} creates multileaved result lists following the \ac{TDM} procedure the probability $P(\mathbf{m} = \textit{BA})$ is twice as high as $P(\mathbf{m} = \textit{AB})$. As a consequence, the expected preference is biased against the first ranker:
\begin{align}
E[P_{12}] < 0. \nonumber
\end{align}
Hence, \ac{SOSM} does not have \emph{fidelity}. This outcome seems to stem from a disconnect between how multileaved results lists are created and how preferences are inferred.

\smallskip\noindent%
To conclude this section, Table~\ref{tab:properties} provides an overview of our findings thus far, i.e., the theoretical requirements that each multileaved comparison method satisfies; we have also included \acs{PPM}, the multileaved comparison method that we will introduce below.

\begin{table}[tb]
\caption{Overview of multileaved comparison methods and whether they meet the \emph{considerateness} and \emph{fidelity} requirements.}
\centering
\begin{tabular}{ l  c c c   }
\toprule
 & \textbf{Considerateness} & \textbf{Fidelity} & \textbf{Source} \\
\midrule
\acs{TDM} & \checkmark & & \citep{Schuth2014a} \\
\acs{OM} & \checkmark & & \citep{Schuth2014a} \\
\acs{PM} & & \checkmark & \citep{schuth2015probabilistic} \\
\acs{SOSM} & \checkmark & & \citep{brost2016improved} \\
\acs{PPM} & \checkmark & \checkmark & this paper \\
\bottomrule 
\end{tabular}
\label{tab:properties}
\end{table}


\section{A Novel Multileaved Comparison Method}
\label{sec:novelmethod}

The previously described multileaved comparison methods are based around direct credit assignment, i.e., credit functions are based on single documents.
In contrast, we introduce a method that estimates differences based on pairwise document preferences.
We prove that this novel method is the only multileaved comparison method that meets the \emph{considerateness} and \emph{fidelity} requirements set out in Section~\ref{sec:multileaving}.

\label{sec:preferencemultileave}
The multileaved comparison method that we introduce is \acf{PPM}. It infers pairwise preferences between documents from clicks and bases comparisons on the agreement of rankers with the inferred preferences.
\ac{PPM} is based on the assumption that a clicked document is preferred to: (a) all of the unclicked documents above it; (b) the next unclicked document. These assumptions are long-established \cite{joachims2002unbiased} and form the basis of pairwise \ac{LTR} \cite{Joachims2002}.

We write $\mathbf{c}_{r(d_i,\mathbf{m})}$ for a click on document $d_i$ displayed in multileaved result list $\mathbf{m}$ at the rank $r(d_i,\mathbf{m})$. For a document pair $(d_i, d_j)$, a click $\mathbf{c}_{r(d_i,\mathbf{m})}$ infers a preference as follows:
\begin{equation}
\begin{aligned}
&c_{r(d_i,\mathbf{m})} \land \neg c_{r(d_j,\mathbf{m})}
\land
\big(\exists i, (c_i \land r(d_j,\mathbf{m}) < i) \lor c_{r(d_j,\mathbf{m}) -1}\big)
\\
&
\qquad {} \Leftrightarrow d_i >_{c} d_j.
\end{aligned}
\end{equation}
In addition, the preference of a ranker $\mathbf{r}$ is denoted by $d_i >_{\mathbf{r}} d_j$.
Pairwise preferences also form the basis for \ac{PBI} introduced by~\citet{he2009evaluation}. However, previous work has shown that \ac{PBI} does not meet the \emph{fidelity} requirements~\citep{hofmann2013fidelity}. Therefore, we do not use \ac{PBI} as a starting point for \ac{PPM}. Instead, \ac{PPM} is derived directly from the \emph{considerateness} and \emph{fidelity} requirements. Consequently, \ac{PPM} constructs multileaved result lists inherently differently and its inference method has \emph{fidelity}, in contrast with \ac{PBI}.

\begin{algorithm}[t]
\begin{algorithmic}[1]
\STATE \textbf{Input}: set of rankers $\mathcal{R}$, rankings $\{\mathbf{l}\}$, documents $D$.
\STATE $\mathbf{m} \leftarrow []$ \hfill \textit{\small // initialize empty multileaving}
\FOR{$n = 1,\ldots,|D|$}
    \STATE $\hat{\Omega}_n \leftarrow \Omega(n,\mathcal{R},D) \setminus \mathbf{m}$  \hfill \textit{\small // choice set of remaining documents} \label{alg:ppm:docset}
    \STATE $d \leftarrow \textit{uniform\_sample}(\hat{\Omega}_n)$   \hfill \textit{\small // uniformly sample next document} \label{alg:ppm:next}
    \STATE $\mathbf{m} \leftarrow \textit{append}(\mathbf{m},d)$   \hfill \textit{\small // add sampled document to multileaving} \label{alg:ppm:append}
\ENDFOR
\RETURN $\mathbf{m}$
\end{algorithmic}
\caption{Multileaved result list construction for \ac{PPM}.}
\label{alg:ppm-listconstruction}
\end{algorithm}

\begin{algorithm}[t]
\begin{algorithmic}[1]
\STATE \textbf{Input}: rankers $\mathcal{R}$, rankings $\{\mathbf{l}\}$, documents $D$, multileaved result list $\mathbf{m}$, clicks $\mathbf{c}$.
\STATE  $P \leftarrow \mathbf{0}$ \hfill \textit{\small // preference matrix of $|\mathcal{R}|\times|\mathcal{R}|$ } 
\FOR{$(d_i, d_j) \in \{(d_i,d_j)\mid d_i >_{\mathbf{c}} d_j\}$}
       \IF{$\ubar{r}(i,j, \mathbf{m}) \geq \bar{r}(i,j)$}
           \STATE ${w} \leftarrow 1$ \hfill \textit{\small // variable to store $P(\ubar{r}(i,j, \mathbf{m})  \geq \bar{r}(i,j))$ }
           \STATE $\textit{min\_x} \leftarrow \min_{d \in \{d_i,d_j\}} \min_{\mathbf{r}_n \in \mathcal{R}} r(d,\mathbf{l}_n)$
           \FOR{$x = \textit{min\_x},\ldots,\bar{r}(i,j)-1$}
               \STATE ${w} \leftarrow {w} \cdot (1 - (|\Omega(x,\mathcal{R},D)|-x-1)^{-1})$
           \ENDFOR
       \FOR{$n = 1,\ldots,|R|$}
       	   \FOR{$m = 1,\ldots,|R|$}
                    \IF{$d_i >_{\mathbf{r}_n} d_j \land n \not = m$}
                        \STATE $P_{nm} \leftarrow P_{nm} + {w}^{-1}$\hfill \textit{\small // result of scoring function $\phi$ }
                    \ELSIF{$n \not = m$}
                         \STATE $P_{nm} \leftarrow P_{nm} - {w}^{-1}$
        	       	   \ENDIF
            \ENDFOR
        \ENDFOR
   \ENDIF   
\ENDFOR
\RETURN $P$
\end{algorithmic}
\caption{Preference inference for \ac{PPM}.}
\label{alg:ppm-inference}
\end{algorithm}

When constructing a multileaved result list $\mathbf{m}$ we want to be able to infer unbiased preferences while simultaneously being \emph{considerate}. Thus, with the requirement for \emph{considerateness} in mind we define a choice set as:
\begin{align}
\Omega(i,\mathcal{R},D) = \{ d \mid  d \in D \land \exists \mathbf{r}_j \in \mathcal{R}, r(d,\mathbf{l}_j) \leq i \}.
\end{align}
This definition is chosen so that any document in $\Omega(i,\mathcal{R}, D)$ can be placed at rank $i$ without breaking the \emph{obstruction} requirement (Equation~\ref{eq:obstruction}). The multileaving method of \ac{PPM} is described in Algorithm~\ref{alg:ppm-listconstruction}. The approach is straightforward: at each rank $n$ the set of documents $\hat{\Omega}_n$ is determined (Line~\ref{alg:ppm:docset}). This set of documents is $\Omega(n,\mathcal{R},D)$ with the previously added documents removed to avoid document repetition. Then, the next document is sampled uniformly from $\hat{\Omega}_n$  (Line~\ref{alg:ppm:next}), thus every document in $\hat{\Omega}_n$ has a probability:
\begin{align}
\frac{1}{|\Omega(n,\mathcal{R},D)| - n + 1}
\end{align}
of being placed at position $n$ (Line~\ref{alg:ppm:append}). Since $\hat{\Omega}_n \subseteq \Omega(n,\mathcal{R},D)$ the resulting $\mathbf{m}$ is guaranteed to be \emph{considerate}.

While the multileaved result list creation method used by \ac{PPM} is simple, its preference inference method is more complicated as it has to meet the \emph{fidelity} requirements. First, the preference found between a ranker $\mathbf{r}_n$ and $\mathbf{r}_m$ from a single interaction $\mathbf{c}$ is determined by:
\begin{align}
P_{nm} = \sum_{d_i >_\mathbf{c} d_j} \phi(d_i,d_j,\mathbf{r}_n, \mathbf{m}, \mathcal{R}) - \phi(d_i,d_j,\mathbf{r}_m, \mathbf{m}, \mathcal{R}),
\end{align}
which sums over all document pairs $(d_i,d_j)$ where interaction $\mathbf{c}$ inferred a preference. Before the scoring function $\phi$ can be defined we introduce the following function:
\begin{align}
 \bar{r}(i,j,\mathcal{R}) = \max_{d \in \{d_i,d_j\}} \min_{\mathbf{r}_n \in \mathcal{R}} r(d,\mathbf{l}_n).
\end{align}
For succinctness we will note $\bar{r}(i,j) =  \bar{r}(i,j,\mathcal{R})$.
Here, $\bar{r}(i,j)$ provides the highest rank at which both documents $d_i$ and $d_j$ can appear in $\mathbf{m}$.
Position $\bar{r}(i,j)$ is important to the document pair $(d_i,d_j)$, since if both documents are in the remaining documents $\hat{\Omega}_{\bar{r}(i,j)}$, then the rest of the multileaved result list creation process is identical for both. To keep notation short we introduce:
\begin{align}
\ubar{r}(i,j, \mathbf{m}) = \min_{d \in \{d_i,d_j\}} r(d,\mathbf{m}).
\end{align}
Therefore, if $\ubar{r}(i,j, \mathbf{m})  \geq \bar{r}(i,j)$ then both documents appear below $\bar{r}(i,j)$. This, in turn, means that both documents are equally likely to appear at any rank:
\begin{equation}
\begin{aligned}
&\forall n,
\phantom{n}
P(\mathbf{m}_n = d_i \mid \ubar{r}(i,j, \mathbf{m})  \geq \bar{r}(i,j))
\\
&
\qquad{}= P(\mathbf{m}_n = d_j \mid  \ubar{r}(i,j, \mathbf{m})  \geq \bar{r}(i,j)).
\end{aligned}
\label{eq:uniformpos}
\end{equation}
The scoring function $\phi$ is then defined as follows:
\begin{align}
 \phi(d_i,d_j,\mathbf{r}, \mathbf{m}) = 
 \begin{cases}
 0, &\ubar{r}(i,j, \mathbf{m}) < \bar{r}(i,j)
 \\
 \frac{-1}{P(\ubar{r}(i,j, \mathbf{m})  \geq \bar{r}(i,j))}, & d_i <_\mathbf{r} d_j
 \\
 \frac{1}{P(\ubar{r}(i,j, \mathbf{m})  \geq \bar{r}(i,j))}, & d_i >_\mathbf{r} d_j,
 \end{cases}
 \label{eq:scoringfunction}
\end{align}
indicating that a zero score is given if one of the documents appears above $\bar{r}(i,j)$.
Otherwise, the value of $\phi$ is positive or negative depending on whether the ranker $\mathbf{r}$ agrees with the inferred preference between $d_i$ and $d_j$. Furthermore, this score is inversely weighed by the probability $P(\ubar{r}(i,j, \mathbf{m})  \geq \bar{r}(i,j))$. Therefore, pairs that are less likely to appear below their threshold $\bar{r}(i,j)$ result in a higher score than for more commonly occuring pairs. Algorithm~\ref{alg:ppm-inference} displays how the inference of \ac{PPM} can be computed. The scoring function $\phi$ was carefully chosen to guarantee \emph{fidelity}, the remainder of this section will sketch the proof for \ac{PPM} meeting its requirements. 

The two requirements for \emph{fidelity} will be discussed in order:

\subsubsection*{Requirement~\ref{fidelity:unbias}} The first fidelity requirement states that under uncorrelated clicks the expected outcome should be zero. Consider the expected preference:
\begin{equation}
\begin{aligned}
E[P_{nm}]
= {} &\sum_{d_i,d_j} \sum_{\mathbf{m}}  P(d_i >_c d_j \mid  \mathbf{m}) P(\mathbf{m}) \\
 &\qquad (\phi(d_i,d_j,\mathbf{r}_n, \mathbf{m}) -  \phi(d_i,d_j,\mathbf{r}_m, \mathbf{m})).
\end{aligned}
\label{eq:expectedoutcome}
\end{equation}
To see that $E[P_{nm}] = 0$ under uncorrelated clicks, take any multileaving $\mathbf{m}$ where $P(\mathbf{m}) > 0$ and $\phi(d_i,d_j,\mathbf{r}, \mathbf{m}) \not= 0$ with $\mathbf{m}_x = d_i$ and $\mathbf{m}_y = d_j$.  Then there is always a multileaved result list $\mathbf{m}'$ that is identical expect for swapping the two documents so that $\mathbf{m}'_x = d_j$ and $\mathbf{m}'_y = d_i$.
The scoring function only gives non-zero values if both documents appear below the threshold $\ubar{r}(i,j, \mathbf{m}) < \bar{r}(i,j)$ (Equation~\ref{eq:scoringfunction}). At this point the probability of each document appearing at any position is the same (Equation~\ref{eq:uniformpos}), thus the following holds:
\begin{align}
P(\mathbf{m}) &= P(\mathbf{m}'),\\
\phi(d_i,d_j,\mathbf{r}_n, \mathbf{m})  &= -\phi(d_j,d_i,\mathbf{r}_n, \mathbf{m}').
\end{align}
Finally, from the definition of uncorrelated clicks (Equation~\ref{eq:uncorrelated}) the following holds:
\begin{align}
P(d_i >_c d_j \mid  \mathbf{m}) &= P(d_j >_c d_i \mid  \mathbf{m}').
\end{align}
As a result, any document pair $(d_i,d_j)$ and multileaving $\mathbf{m}$ that affects the expected outcome is cancelled by the multileaving $\mathbf{m}'$. Therefore, we can conclude that $E[P_{nm}] = 0$ under uncorrelated clicks, and that \ac{PPM} meets the first requirement of \emph{fidelity}.

\subsubsection*{Requirement~\ref{fidelity:pareto}} The second \emph{fidelity} requirement states that under correlated clicks a ranker that Pareto dominates all other rankers should win the multileaved comparison. Therefore, the expected value for a Pareto dominating ranker $r_n$ should be:
\begin{align}
\forall m, n \not = m \rightarrow E[P_{nm}] > 0.
\end{align}
Take any other ranker $\mathbf{r}_m$ that is thus Pareto dominated by $\mathbf{r}_n$.
The proof for the first requirement shows that $E[P_{nm}]$ is not affected by any pair of documents $d_i,d_j$  with the same relevance label.
Furthermore, any pair on which $\mathbf{r}_n$ and $\mathbf{r}_m$ agree will not affect the expected outcome since:
\begin{equation}
\begin{aligned}
&(d_i >_{\mathbf{r}_n} d_j \leftrightarrow d_i >_{\mathbf{r}_m} d_j) \Rightarrow{}\\
&\qquad \phi(d_i,d_j,\mathbf{r}_n, \mathbf{m}) -  \phi(d_i,d_j,\mathbf{r}_m, \mathbf{m}) = 0.
\end{aligned}
\end{equation}
Then, for any relevant document $d_i$, consider the set of documents that $\mathbf{r}_n$ incorrectly prefers over $d_i$:
\begin{align}
A = \{d_j \mid  \neg \mathit{rel}(d_j) \land d_j >_{\mathbf{r}_n} d_i \}
\end{align}
and the set of documents that $\mathbf{r}_m$ incorrectly prefers over $d_i$ and places higher than where $\mathbf{r}_n$ places $d_i$:
\begin{align}
B = \{d_j \mid  \neg \mathit{rel}(d_j) \land  d_j >_{\mathbf{r}_m} d_i \land r(d_j, \mathbf{l}_m) <  r(d_i, \mathbf{l}_n)\}.
\end{align}
Since $\mathbf{r}_n$ Pareto dominates $\mathbf{r}_m$, it has the same or fewer incorrect preferences: $|A| \leq |B|$. Furthermore, for any document $d_j$ in either $A$ or $B$ the threshold of the pair $d_i,d_j$ is the same:
\begin{align}
\forall d_j \in A \cup B, \bar{r}(i,j) = r(d_i, \mathbf{l}_n).
\end{align}
Therefore, all pairs with documents from $A$ and $B$ will only get a non-zero value from $\phi$ if they both appear at or below $r(d_i, \mathbf{l}_n)$.
Then using  Equation~\ref{eq:uniformpos} and the Bayes rule we see:
\begin{equation}
\begin{aligned}
\forall (d_j, d_l) \in  A \cup B,
&
\frac{
P(\mathbf{m}_x = d_j, \ubar{r}(i,j, \mathbf{m})  \geq \bar{r}(i,j,\mathcal{R}))
}{
P(\ubar{r}(i,j, \mathbf{m})  \geq \bar{r}(i,j,\mathcal{R}))
}
\\
&{}=
\frac{
P(\mathbf{m}_x = d_l, \ubar{r}(i,l, \mathbf{m})  \geq \bar{r}(i,l,\mathcal{R}))
}{
P(\ubar{r}(i,l, \mathbf{m})  \geq \bar{r}(i,l,\mathcal{R}))
}.
\end{aligned}
\end{equation}
Similarly, the reweighing of $\phi$ ensures that every pair in $A$ and $B$ contributes the same to the expected outcome. Thus, if both rankers rank $d_i$ at the same position the following sum:
\begin{equation}
\begin{aligned}
& \sum_{d_j \in A \cup B} \sum_{\mathbf{m}} P(\mathbf{m}) \cdot{}\\
&\qquad \left[
 P(d_i >_c d_j \mid  \mathbf{m}) 
 (\phi(d_i,d_j,\mathbf{r}_n, \mathbf{m}) -  \phi(d_i,d_j,\mathbf{r}_m, \mathbf{m}))\right.
 \\
&\qquad \left.{}+ P(d_j >_c d_i \mid  \mathbf{m})
 (\phi(d_j,d_i,\mathbf{r}_n, \mathbf{m}) -  \phi(d_j,d_i,\mathbf{r}_m, \mathbf{m}))\right]
 \end{aligned}
 \end{equation}
 will be zero if $|A| = |B|$ and positive if $|A| < |B|$ under correlated clicks. Moreover, since $\mathbf{r}_n$ Pareto dominates $\mathbf{r}_m$, there will be at least one document $d_j$ where:
 \begin{align}
 \exists d_i, \exists d_j, \mathit{rel}(d_i) \land \neg \mathit{rel}(d_j) \land r(d_i, \mathbf{l}_n) = r(d_j, \mathbf{l}_m).
 \end{align}
This means that the expected outcome (Equation~\ref{eq:expectedoutcome}) will always be positive under correlated clicks, i.e., $E[P_{nm}] > 0$, for a Pareto dominating ranker $\mathbf{r}_n$ and any other ranker $\mathbf{r}_m$.

In summary, we have introduced a new multileaved comparison method, \ac{PPM}, which we have shown to be \emph{considerate} and to have \emph{fidelity}. We further note that \ac{PPM} has polynomial complexity: to calculate $P(\ubar{r}(i,j, \mathbf{m})  \geq \bar{r}(i,j))$ only the size of the choice sets $\Omega$ and the first positions at which $d_i$ and $d_j$ occur in $\Omega$ have to be known.

\section{Experiments}
\label{sec:experiments}

In order to answer Research Question~\ref{rq:sensitive} posed in Section~\ref{sec:intro} several experiments were performed to evaluate the \emph{sensitivity} of \ac{PPM}. The methodology of evaluation follows previous work on interleaved and multileaved comparison methods \cite{Schuth2014a, hofmann2011probabilistic, schuth2015probabilistic, hofmann2013fidelity, brost2016improved} and is completely reproducible.\footnote{\url{https://github.com/HarrieO/PairwisePreferenceMultileave}}

\subsection{Ranker selection and comparisons}
In order to make fair comparisons between rankers, we will use the \ac{OLTR} datasets described in Section~\ref{sec:datasets}. From the feature representations in these datasets a handpicked set of features was taken and used as ranking models. To match the real-world scenario as best as possible this selection consists of features which are known to perform well as relevance signals independently. This selection includes but is not limited to: BM25, LMIR.JM, Sitemap, PageRank, HITS and TF.IDF.

Then the ground-truth comparisons between the rankers are based on their NDCG scores computed on a held-out test set, resulting in a binary preference matrix $P_{nm}$ for all ranker pairs $(\mathbf{r}_n,\mathbf{r}_m)$:
\begin{align}
P_{nm} = NDCG(\mathbf{r}_n) - NDCG(\mathbf{r}_m).
\end{align}
The metric by which multileaved comparison methods are compared is the \emph{binary error}, $E_{bin}$~ \citep{Schuth2014a, brost2016improved, schuth2015probabilistic}. Let $\hat{P}_{nm}$ be the preference inferred by a multileaved comparison method; then the error is:
\begin{align}
E_{bin} = \frac{\sum_{n,m\in\mathcal{R} \land n \not = m}\sgn(\hat{P}_{nm}) \not = sgn(P_{nm})}{|\mathcal{R}| \times (|\mathcal{R}| - 1)}.
\end{align}

\subsection{Datasets}
\label{sec:datasets}

Our experiments are performed over ten publicly available \acs{OLTR} datasets with varying sizes and representing different search tasks. Each dataset consists of a set of queries and a set of corresponding documents for every query. While queries are represented only by their identifiers, feature representations and relevance labels are available for every document-query pair. Relevance labels are graded differently by the datasets depending on the task they model, for instance, navigational datasets have binary labels for not relevant (0), and relevant (1), whereas  most informational tasks have labels ranging from not relevant (0), to perfect relevancy (5).
Every dataset consists of five folds, each dividing the dataset in different training, validation and test partitions.

The first publicly available \emph{Learning to Rank} datasets are distributed as LETOR 3.0 and 4.0~\cite{letor}; they use representations of 45, 46, or 64 features encoding ranking models such as TF.IDF, BM25, Language Modelling, PageRank, and HITS on different parts of the documents. The datasets in LETOR are divided by their tasks, most of which come from the TREC Web Tracks between 2003 and 2008 \cite{craswell2003overview,clarke2009overview}.
\emph{HP2003, HP2004, NP2003, NP2004, TD2003} and \emph{TD2004} each contain between 50 and 150 queries and 1,000 judged documents per query and use binary relevance labels. Due to their similarity we report average results over these six datasets noted as \emph{LETOR 3.0}.
The \emph{OH\-SU\-MED} dataset is based on the query log of the search engine on the MedLine abstract database, and contains 106 queries. The last two datasets, \emph{MQ2007} and \emph{MQ2008}, were based on the Million Query Track \cite{allan2007million} and consist of 1,700 and 800 queries, respectively, but have far fewer assessed documents per query.

The \emph{MLSR-WEB10K} dataset \cite{Qin2013Letor} consists of 10,000 queries obtained from a retired labelling set of a commercial web search engine. 
The datasets uses 136 features to represent its documents, each query has around 125 assessed documents.

Finally, we note there are more \emph{OLTR} datasets available \cite{Chapelle2011, dato2016fast}, but there is no public information about their feature representations. Therefore, they are unfit for our evaluation as no selection of well performing ranking features can be made.

\subsection{Simulating user behavior}
\label{sec:experiments:users}

\begin{table}[tb]
\caption{Instantiations of Cascading Click Models~\cite{guo09:efficient} as used for simulating user behaviour in experiments.}
\label{tab:clickmodels}
\centering
\begin{tabular}{ @{} l c c c c c c c c c c }
\toprule
& \multicolumn{5}{c}{\small $P(\mathit{click}=1\mid R)$} & \multicolumn{5}{c}{\small $P(\mathit{stop}=1\mid R)$} \\
\cmidrule(r){2-6}\cmidrule(l){7-11}
\small $R$ & \small \emph{$ 0$} & \small \emph{$ 1$}  & \small \emph{$ 2$} & \small \emph{$ 3$} & \small \emph{$ 4$}
 & \small \emph{$0$} & \small \emph{$ 1$} & \small \emph{$ 2$} & \small \emph{$ 3$} & \small \emph{$ 4$} \\
\midrule
\small \emph{perf} & \small 0.0 & \small 0.2 & \small 0.4 & \small 0.8 & \small 1.0 & \small 0.0 & \small 0.0 & \small 0.0 & \small 0.0 & \small 0.0 \\
\small \emph{nav} & \small ~~0.05 & \small 0.3 & \small 0.5 & \small 0.7 & \small ~~0.95 & \small 0.2 & \small 0.3 & \small 0.5 & \small 0.7 & \small 0.9 \\
\small \emph{inf} & \small 0.4 & \small 0.6 & \small 0.7 & \small 0.8 & \small 0.9 & \small 0.1 & \small 0.2 & \small 0.3 & \small 0.4 & \small 0.5 \\
\bottomrule
\end{tabular}
\vspace*{-\baselineskip}
\end{table}

While experiments using real users are preferred \cite{chuklin2015comparative, chapelle2012large, kharitonov2015generalized, yue2010learning}, most researchers do not have access to search engines. As a result the most common way of comparing online evaluation methods is by using simulated user behaviour~\citep{Schuth2014a, hofmann2011probabilistic, schuth2015probabilistic, hofmann2013fidelity, brost2016improved}. Such simulated experiments show the performance of multileaved comparison methods when user behaviour adheres to a few simple assumptions. 

Our experiments follow the precedent set by previous work on online evaluation:
First, a user issues a query simulated by uniformly sampling a query from the static dataset. Subsequently, the multileaved comparison method constructs the multileaved result list of documents to display. The behavior of the user after receiving this list is simulated using a \emph{cascade click model}~\cite{chuklin-click-2015,guo09:efficient}. This model assumes a user to examine documents in their displayed order. For each document that is considered the user decides whether it warrants a click, which is modeled as the conditional probability $P(click=1\mid R)$ where $R$ is the relevance label provided by the dataset. Accordingly, \emph{cascade click model} instantiations increase the probability of a click with the degree of the relevance label. After the user has clicked on a document their information need may be satisfied; otherwise they continue considering the remaining documents. The probability of the user not examining more documents after clicking is modeled as $P(stop=1\mid R)$,
where it is more likely that the user is satisfied from a very relevant document. At each impression we display $k=10$ documents to the user.

Table~\ref{tab:clickmodels} lists the three instantiations of cascade click models that we use for this paper.
The first models a \emph{perfect} user (\emph{perf}) who considers every document and clicks on all relevant documents and nothing else.
Secondly, the \emph{navigational} instantiation (\emph{nav}) models a user performing a navigational task who is mostly looking for a single highly relevant document. Finally, the \emph{informational} instantiation (\emph{inf}) models a user without a very specific information need who typically clicks on multiple documents.
These three models have increasing levels of noise, as the behavior of each depends less on the relevance labels of the displayed documents.

\subsection{Experimental runs}
\label{sec:experiments:runs}

Each experimental run consists of applying a multileaved comparison method to a sequence of $T=10,000$ simulated user impressions. To see the effect of the number of rankers in a comparison, our runs consider $|\mathcal{R}| = 5$, $|\mathcal{R}| = 15$, and $|\mathcal{R}| = 40$. However only the \emph{MSLR} dataset contains $|\mathcal{R}| = 40$ rankers. Every run is repeated for every click model to see how different behaviours affect performance. For statistical significance every run is repeated 25 times per fold, which means that 125 runs are conducted for every dataset and click model pair. Since our evaluation covers five multileaved comparison methods, we generate over 393 million impressions in total. We test for statistical significant differences using a two tailed t-test. Note that the results reported on the LETOR 3.0 data are averaged over six datasets and thus span 750 runs per datapoint.

The parameters of the baselines are selected based on previous work on the same datasets; for \ac{OM} the sample size $\eta=10$ was chosen as reported by \citet{Schuth2014a}; for \ac{PM} the degree $\tau=3.0$ was chosen according to \citet{hofmann2011probabilistic} and the sample size $\eta=10,000$ in accordance with \citet{schuth2015probabilistic}.


\begin{figure}[tb]
\centering
\includegraphics[width=\columnwidth]{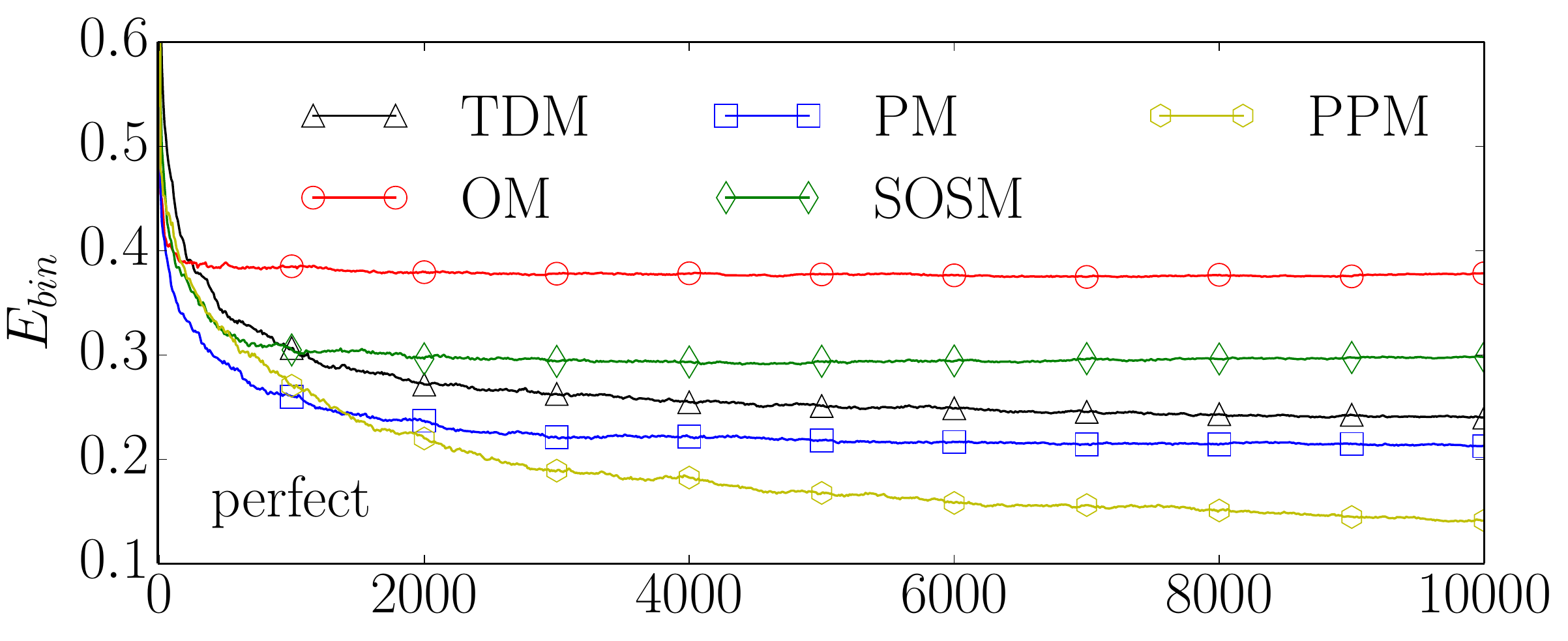}
\includegraphics[width=\columnwidth]{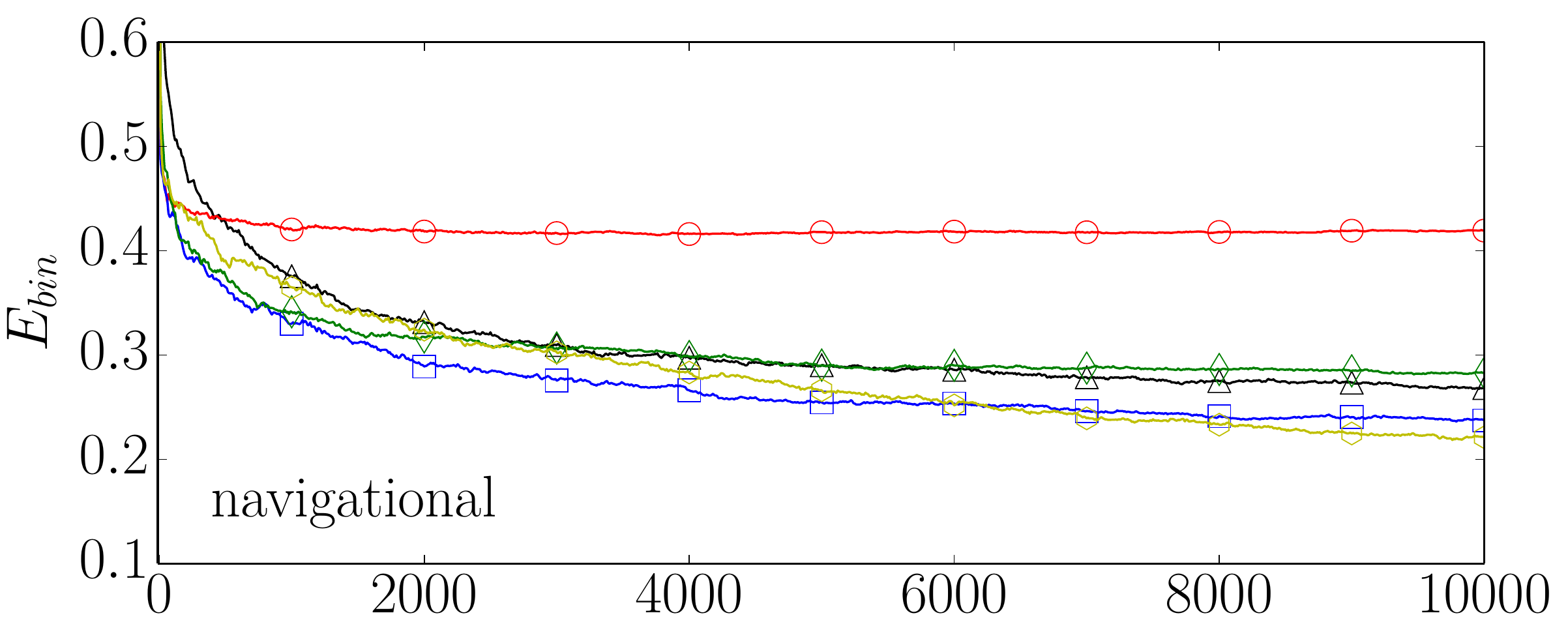}
\includegraphics[width=\columnwidth]{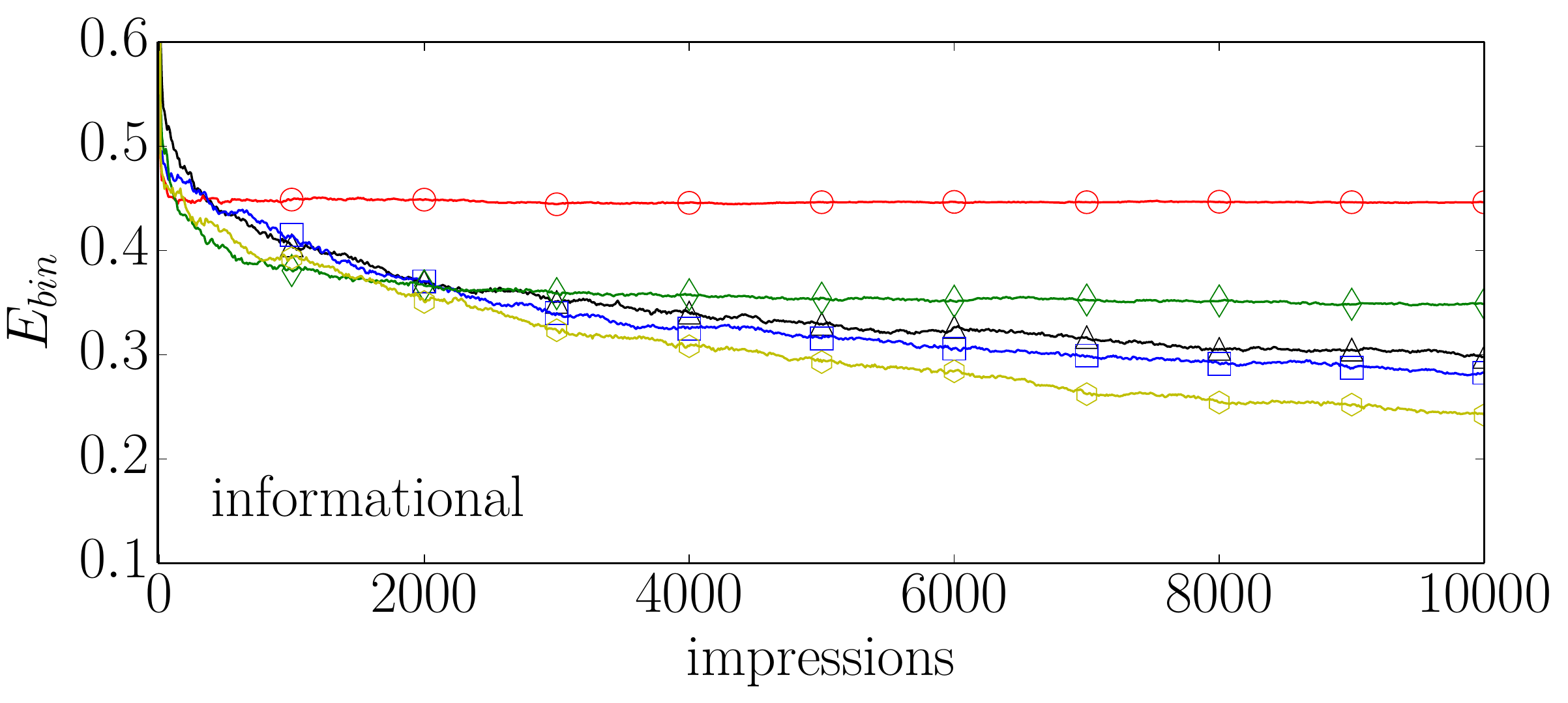}
\caption{The binary error of different multileaved comparison methods on comparisons of $|\mathcal{R}| = 15$ rankers on the \emph{MSLR-WEB10k} dataset.}
\label{fig:binaryerror}

\end{figure}

\section{Results and Analysis}
\label{sec:results}

We answer Research Question~\ref{rq:sensitive} by evaluating the \emph{sensitivity} of \ac{PPM} based on the results of the experiments detailed in Section~\ref{sec:experiments}.

The results of the experiments with a smaller number of rankers: $|\mathcal{R}| = 5$ are displayed in Table~\ref{tab:5rankers}. Here we see that after 10,000 impressions \ac{PPM} has a significantly lower error on many datasets and at all levels of interaction noise. Furthermore, for $|\mathcal{R}| = 5$ there are no significant losses in performance under any circumstances.

When $|\mathcal{R}| = 15$ as displayed in Table~\ref{tab:15rankers}, we see a single case where \ac{PPM} performs worse than a previous method: on \emph{MQ2007} under the \emph{perfect} click model \ac{SOSM} performs significantly better than \ac{PPM}. However, on the same dataset \ac{PPM} performs significantly better under the \emph{informational} click model. Furthermore, there are more significant improvements for $|\mathcal{R}| = 15$ than when the number of rankers is the smaller $|\mathcal{R}| = 5$.

Finally, when the number of rankers in the comparison is increased to $|\mathcal{R}| = 40$ as displayed in Table~\ref{tab:40rankers}, \ac{PPM} still provides significant improvements.

We conclude that \ac{PPM} always provides a performance that is at least as good as any existing method. Moreover, \ac{PPM} is robust to noise as we see more significant improvements under click-models with increased noise. Furthermore, since improvements are found with the number of rankers $|\mathcal{R}|$ varying from $5$ to $40$, we conclude that \ac{PPM} is scalable in the comparison size.
Additionally, the dataset type seems to affect the relative performance of the methods. For instance, on \emph{LETOR 3.0} little significant differences are found, whereas the \emph{MSLR} dataset displays the most significant improvements. This suggests that on more artificial data, i.e., the smaller datasets simulating navigational tasks, the differences are fewer, while on the other hand on large commercial data the preference for \ac{PPM} increases further.
Lastly, Figure~\ref{fig:binaryerror} displays the binary error of all multileaved comparison methods on the \emph{MSLR} dataset over 10,000 impressions. Under the \emph{perfect} click model we see that all of the previous methods display converging behavior around 3,000 impressions. In contrast, the error of \ac{PPM} continues to drop throughout the experiment. The fact that the existing methods converge at a certain level of error in the absence of click-noise is indicative that they are lacking in \emph{sensitivity}.

\begin{table*}[tb]
\centering
\caption{
The binary error $E_{bin}$ of all multileaved comparison methods after 10,000 impressions on comparisons of $|\mathcal{R}| = 5$ rankers. Average per dataset and click model; standard deviation in brackets. The best performance per click model and dataset is noted in bold, statistically significant improvements of \acs{PPM} are noted by \dubbelneer $(p < 0.01)$ and \enkelneer $(p < 0.05)$ and losses by \dubbelop $ $ and \enkelop $ $ respectively or $\circ$ for no difference, per baseline.
}

\begin{tabularx}{\textwidth}{ X  X X X X l @{~}c@{}c@{}c@{}c }
\toprule
 & \multicolumn{1}{l}{ \small \textbf{\acs{TDM}}}  & \multicolumn{1}{l}{ \small \textbf{\acs{OM}}}  & \multicolumn{1}{l}{ \small \textbf{\acs{PM}}}  & \multicolumn{1}{l}{ \small \textbf{SOSM}}  & \multicolumn{1}{l}{ \small \textbf{\acs{PPM}}} \\
\midrule
& \multicolumn{5}{c}{\textit{perfect}} \\
\midrule
LETOR 3.0 & {\small 0.16 {\tiny( 0.13)}} & \textbf {\small 0.14 {\tiny( 0.15)}} & {\small 0.15 {\tiny( 0.15)}} & {\small 0.16 {\tiny( 0.15)}} & \textbf {\small 0.14 {\tiny( 0.13)}} &{$\circ$}& {$\circ$} & {$\circ$} & {$\circ$} 
\\
MQ2007 & {\small 0.19 {\tiny( 0.16)}} & {\small 0.22 {\tiny( 0.18)}} & \textbf {\small 0.16 {\tiny( 0.14)}} & {\small 0.18 {\tiny( 0.16)}} & \textbf {\small 0.16 {\tiny( 0.14)}} & {$\circ$} & {\tiny \dubbelneer} & {$\circ$} & {$\circ$} 
\\
MQ2008 & {\small 0.15 {\tiny( 0.12)}} & {\small 0.19 {\tiny( 0.14)}} & {\small 0.16 {\tiny( 0.12)}} & {\small 0.18 {\tiny( 0.15)}} & \textbf {\small 0.14 {\tiny( 0.12)}} & {$\circ$} & {\tiny \dubbelneer} & {$\circ$} & {\tiny \enkelneer} 
\\
MSLR-WEB10k & {\small 0.23 {\tiny( 0.13)}} & {\small 0.27 {\tiny( 0.17)}} & {\small 0.20 {\tiny( 0.14)}} & {\small 0.25 {\tiny( 0.18)}} & \textbf {\small 0.14 {\tiny( 0.13)}} & {\tiny \dubbelneer} & {\tiny \dubbelneer} & {\tiny \dubbelneer} & {\tiny \dubbelneer} 
\\
OHSUMED & {\small 0.14 {\tiny( 0.12)}} & {\small 0.19 {\tiny( 0.15)}} & \textbf {\small 0.11 {\tiny( 0.09)}} & \textbf {\small 0.11 {\tiny( 0.10)}} & \textbf {\small 0.11 {\tiny( 0.10)}} & {\tiny \dubbelneer} & {\tiny \dubbelneer} & {$\circ$} & {$\circ$} 
\\
\midrule
& \multicolumn{5}{c}{\textit{navigational}} \\
\midrule
LETOR 3.0 & {\small 0.16 {\tiny( 0.13)}} & \textbf {\small 0.15 {\tiny( 0.15)}} & \textbf {\small 0.15 {\tiny( 0.14)}} & {\small 0.17 {\tiny( 0.15)}} & {\small 0.16 {\tiny( 0.14)}} & {$\circ$} & {$\circ$} & {$\circ$} & {$\circ$} 
\\
MQ2007 & {\small 0.21 {\tiny( 0.17)}} & {\small 0.33 {\tiny( 0.21)}} & {\small 0.18 {\tiny( 0.12)}} & {\small 0.29 {\tiny( 0.23)}} & \textbf {\small 0.17 {\tiny( 0.14)}} & {$\circ$} & {\tiny \dubbelneer} & {$\circ$} & {\tiny \dubbelneer} 
\\
MQ2008 & {\small 0.17 {\tiny( 0.14)}} & {\small 0.21 {\tiny( 0.20)}} & {\small 0.17 {\tiny( 0.15)}} & {\small 0.23 {\tiny( 0.18)}} & \textbf {\small 0.15 {\tiny( 0.13)}} & {$\circ$} & {\tiny \dubbelneer} & {$\circ$} & {\tiny \dubbelneer} 
\\
MSLR-WEB10k & {\small 0.24 {\tiny( 0.14)}} & {\small 0.32 {\tiny( 0.20)}} & {\small 0.24 {\tiny( 0.17)}} & {\small 0.31 {\tiny( 0.19)}} & \textbf {\small 0.20 {\tiny( 0.15)}} & {\tiny \enkelneer} & {\tiny \dubbelneer} & {\tiny \enkelneer} & {\tiny \dubbelneer} \\
OHSUMED & \textbf {\small 0.12 {\tiny( 0.11)}} & {\small 0.27 {\tiny( 0.19)}} & {\small 0.14 {\tiny( 0.12)}} & {\small 0.23 {\tiny( 0.17)}} & {\small 0.13 {\tiny( 0.12)}} & {$\circ$} & {\tiny \dubbelneer} & {$\circ$} & {\tiny \dubbelneer} 
\\
\midrule
& \multicolumn{5}{c}{\textit{informational}} \\
\midrule
LETOR 3.0 & {\small 0.16 {\tiny( 0.14)}} & {\small 0.22 {\tiny( 0.19)}} & \textbf {\small 0.14 {\tiny( 0.11)}} & {\small 0.17 {\tiny( 0.15)}} & {\small 0.15 {\tiny( 0.13)}} & {$\circ$} & {\tiny \dubbelneer} & {$\circ$} & {$\circ$} 
\\
MQ2007 & {\small 0.23 {\tiny( 0.15)}} & {\small 0.41 {\tiny( 0.26)}} & {\small 0.23 {\tiny( 0.15)}} & {\small 0.37 {\tiny( 0.23)}} & \textbf {\small 0.17 {\tiny( 0.16)}} & {\tiny \dubbelneer} & {\tiny \dubbelneer} & {\tiny \dubbelneer} & {\tiny \dubbelneer} 
\\
MQ2008 & {\small 0.18 {\tiny( 0.13)}} & {\small 0.28 {\tiny( 0.19)}} & {\small 0.18 {\tiny( 0.16)}} & {\small 0.23 {\tiny( 0.18)}} & \textbf {\small 0.17 {\tiny( 0.14)}} & {$\circ$} & {\tiny \dubbelneer} & {$\circ$} & {\tiny \dubbelneer} 
\\
MSLR-WEB10k & {\small 0.27 {\tiny( 0.18)}} & {\small 0.42 {\tiny( 0.23)}} & {\small 0.24 {\tiny( 0.17)}} & {\small 0.36 {\tiny( 0.20)}} & \textbf {\small 0.19 {\tiny( 0.17)}} & {\tiny \dubbelneer} & {\tiny \dubbelneer} & {\tiny \enkelneer} & {\tiny \dubbelneer} 
\\
OHSUMED & {\small 0.13 {\tiny( 0.10)}} & {\small 0.37 {\tiny( 0.24)}} & \textbf {\small 0.12 {\tiny( 0.11)}} & {\small 0.27 {\tiny( 0.21)}} & \textbf {\small 0.12 {\tiny( 0.10)}} & {$\circ$} & {\tiny \dubbelneer} & {$\circ$} & {\tiny \dubbelneer} 
\\
\bottomrule
\end{tabularx}

\vspace{-.5\baselineskip}
\label{tab:5rankers}
\end{table*}

\begin{table*}[tb]
\centering
\caption{
The binary error $E_{bin}$ after 10,000 impressions on comparisons of $|\mathcal{R}| = 15$ rankers. Notation is identical to Table~\ref{tab:5rankers}.
}

\begin{tabularx}{\textwidth}{ X  X X X X l @{~}c@{}c@{}c@{}c }
\toprule
 & \multicolumn{1}{l}{ \small \textbf{\acs{TDM}}}  & \multicolumn{1}{l}{ \small \textbf{\acs{OM}}}  & \multicolumn{1}{l}{ \small \textbf{\acs{PM}}}  & \multicolumn{1}{l}{ \small \textbf{SOSM}}  & \multicolumn{1}{l}{ \small \textbf{\acs{PPM}}} \\
\midrule
& \multicolumn{5}{c}{\textit{perfect}} \\
\midrule
LETOR 3.0 & {\small 0.16 {\tiny( 0.07)}} & \textbf {\small 0.14 {\tiny( 0.08)}} & {\small 0.15 {\tiny( 0.07)}} & {\small 0.17 {\tiny( 0.08)}} & {\small 0.16 {\tiny( 0.08)}} & {$\circ$} & {$\circ$} & {$\circ$} & {$\circ$} 
\\
MQ2007 & {\small 0.20 {\tiny( 0.07)}} & {\small 0.25 {\tiny( 0.09)}} & {\small 0.18 {\tiny( 0.06)}} & \textbf {\small 0.15 {\tiny( 0.07)}} & {\small 0.19 {\tiny( 0.07)}} & {$\circ$} & {\tiny \dubbelneer} & {$\circ$} & {\tiny \dubbelop} 
\\
MQ2008 & {\small 0.16 {\tiny( 0.05)}} & {\small 0.17 {\tiny( 0.05)}} & {\small 0.16 {\tiny( 0.05)}} & \textbf {\small 0.15 {\tiny( 0.07)}} & \textbf {\small 0.15 {\tiny( 0.06)}} & {$\circ$} & {\tiny \enkelneer} & {$\circ$} & {$\circ$} 
\\
MSLR-WEB10k & {\small 0.24 {\tiny( 0.07)}} & {\small 0.38 {\tiny( 0.11)}} & {\small 0.21 {\tiny( 0.06)}} & {\small 0.30 {\tiny( 0.08)}} & \textbf {\small 0.14 {\tiny( 0.05)}} & {\tiny \dubbelneer} & {\tiny \dubbelneer} & {\tiny \dubbelneer} & {\tiny \dubbelneer} 
\\
OHSUMED & {\small 0.14 {\tiny( 0.03)}} & {\small 0.18 {\tiny( 0.05)}} & {\small 0.13 {\tiny( 0.03)}} & {\small 0.13 {\tiny( 0.03)}} & \textbf {\small 0.11 {\tiny( 0.03)}} & {\tiny \dubbelneer} & {\tiny \dubbelneer} & {\tiny \dubbelneer} & {\tiny \dubbelneer} 
\\
\midrule
& \multicolumn{5}{c}{\textit{navigational}} \\
\midrule
LETOR 3.0 & {\small 0.16 {\tiny( 0.08)}} & {\small 0.16 {\tiny( 0.09)}} & \textbf {\small 0.15 {\tiny( 0.08)}} & {\small 0.17 {\tiny( 0.08)}} & {\small 0.17 {\tiny( 0.08)}} & {$\circ$} & {$\circ$} & {$\circ$} & {$\circ$} 
\\
MQ2007 & {\small 0.24 {\tiny( 0.07)}} & {\small 0.33 {\tiny( 0.11)}} & \textbf {\small 0.20 {\tiny( 0.07)}} & {\small 0.22 {\tiny( 0.08)}} & {\small 0.21 {\tiny( 0.08)}} & {\tiny \dubbelneer} & {\tiny \dubbelneer} & {$\circ$} & {$\circ$} 
\\
MQ2008 & {\small 0.19 {\tiny( 0.05)}} & {\small 0.21 {\tiny( 0.07)}} & \textbf {\small 0.16 {\tiny( 0.05)}} & {\small 0.18 {\tiny( 0.06)}} & \textbf {\small 0.16 {\tiny( 0.06)}} & {\tiny \dubbelneer} & {\tiny \dubbelneer} & {$\circ$} & {\tiny \dubbelneer} 
\\
MSLR-WEB10k & {\small 0.27 {\tiny( 0.07)}} & {\small 0.42 {\tiny( 0.12)}} & {\small 0.24 {\tiny( 0.06)}} & {\small 0.28 {\tiny( 0.09)}} & \textbf {\small 0.22 {\tiny( 0.08)}} & {\tiny \dubbelneer} & {\tiny \dubbelneer} & {$\circ$} & {\tiny \dubbelneer} 
\\
OHSUMED & {\small 0.14 {\tiny( 0.04)}} & {\small 0.25 {\tiny( 0.07)}} & \textbf {\small 0.13 {\tiny( 0.03)}} & {\small 0.18 {\tiny( 0.06)}} & \textbf {\small 0.13 {\tiny( 0.04)}} & {$\circ$} & {\tiny \dubbelneer} & {$\circ$} & {\tiny \dubbelneer} 
\\
\midrule
& \multicolumn{5}{c}{\textit{informational}} \\
\midrule
LETOR 3.0 & {\small 0.18 {\tiny( 0.07)}} & {\small 0.20 {\tiny( 0.11)}} & {\small 0.17 {\tiny( 0.08)}} & \textbf {\small 0.16 {\tiny( 0.08)}} & {\small 0.18 {\tiny( 0.08)}} & {$\circ$} & {\tiny \enkelneer} & {$\circ$} & {$\circ$} 
\\
MQ2007 & {\small 0.28 {\tiny( 0.07)}} & {\small 0.42 {\tiny( 0.14)}} & {\small 0.26 {\tiny( 0.08)}} & {\small 0.28 {\tiny( 0.11)}} & \textbf {\small 0.21 {\tiny( 0.08)}} & {\tiny \dubbelneer} & {\tiny \dubbelneer} & {\tiny \dubbelneer} & {\tiny \dubbelneer} 
\\
MQ2008 & {\small 0.23 {\tiny( 0.06)}} & {\small 0.26 {\tiny( 0.11)}} & {\small 0.18 {\tiny( 0.06)}} & {\small 0.20 {\tiny( 0.06)}} & \textbf {\small 0.15 {\tiny( 0.06)}} & {\tiny \dubbelneer} & {\tiny \dubbelneer} & {\tiny \dubbelneer} & {\tiny \dubbelneer} 
\\
MSLR-WEB10k & {\small 0.30 {\tiny( 0.09)}} & {\small 0.45 {\tiny( 0.12)}} & {\small 0.28 {\tiny( 0.08)}} & {\small 0.35 {\tiny( 0.11)}} & \textbf {\small 0.24 {\tiny( 0.08)}} & {\tiny \dubbelneer} & {\tiny \dubbelneer} & {\tiny \dubbelneer} & {\tiny \dubbelneer} 
\\
OHSUMED & {\small 0.15 {\tiny( 0.03)}} & {\small 0.42 {\tiny( 0.09)}} & \textbf {\small 0.13 {\tiny( 0.03)}} & {\small 0.25 {\tiny( 0.06)}} & \textbf {\small 0.13 {\tiny( 0.04)}} & {\tiny \dubbelneer} & {\tiny \dubbelneer} & {$\circ$} & {\tiny \dubbelneer} 
\\
\bottomrule
\end{tabularx}
\label{tab:15rankers}
\vspace{-\baselineskip}
\end{table*}

\begin{table*}[tb]
\centering
\caption{The binary error $E_{bin}$ of all multileaved comparison methods after 10,000 impressions on comparisons of $|\mathcal{R}| = 40$ rankers.
Averaged over the \emph{MSLR-WEB10k}, notation is identical to Table~\ref{tab:5rankers}.
}

\begin{tabularx}{\textwidth}{ X  X X X X l @{~}c@{}c@{}c@{}c }
\toprule
 & \multicolumn{1}{l}{ \small \textbf{\acs{TDM}}}  & \multicolumn{1}{l}{ \small \textbf{\acs{OM}}}  & \multicolumn{1}{l}{ \small \textbf{\acs{PM}}}  & \multicolumn{1}{l}{ \small \textbf{SOSM}}  & \multicolumn{1}{l}{ \small \textbf{\acs{PPM}}} 
 \\
 \midrule
\textit{perfect} & {\small 0.26 {\tiny( 0.03)}} & {\small 0.43 {\tiny( 0.02)}} & {\small 0.23 {\tiny( 0.02)}} & {\small 0.31 {\tiny( 0.02)}} & \textbf {\small 0.18 {\tiny( 0.04)}} & {\tiny \dubbelneer} & {\tiny \dubbelneer} & {\tiny \dubbelneer} & {\tiny \dubbelneer} 
\\
\textit{navigational} & {\small 0.31 {\tiny( 0.03)}} & {\small 0.44 {\tiny( 0.01)}} & {\small 0.25 {\tiny( 0.03)}} & \textbf {\small 0.23 {\tiny( 0.03)}} & {\small 0.24 {\tiny( 0.05)}} & {\tiny \dubbelneer} & {\tiny \dubbelneer} & {\tiny \dubbelneer} & {$\circ$} 
\\
\textit{informational} & {\small 0.37 {\tiny( 0.04)}} & {\small 0.47 {\tiny( 0.01)}} & {\small 0.30 {\tiny( 0.05)}} & {\small 0.34 {\tiny( 0.05)}} & \textbf {\small 0.27 {\tiny( 0.06)}} & {\tiny \dubbelneer} & {\tiny \dubbelneer} & {\tiny \dubbelneer} & {\tiny \dubbelneer} 
\\
\bottomrule
\end{tabularx}
\label{tab:40rankers}
\vspace{-\baselineskip}
\end{table*}

Overall, our results show that \ac{PPM} reaches a lower level of error than previous methods seem to be capable of. This feat can be observed on a diverse set of datasets, various levels of interaction noise and for different comparison sizes. To answer Research Question~\ref{rq:sensitive}: from our results we conclude that \ac{PPM} is more sensitive than any existing multileaved comparison method.


\section{Conclusion}
\label{sec:conclusion}

In this paper we have examined multileaved comparison methods for evaluating ranking models online.

We have presented a new multileaved comparison method, \acf{PPM}, that is more sensitive to user preferences than existing methods. Additionally, we have proposed a theoretical framework for assessing multileaved comparison methods, with \emph{considerateness} and \emph{fidelity} as the two key requirements. We have shown that no method published prior to \ac{PPM} has \emph{fidelity} without lacking \emph{considerateness}. In other words, prior to \ac{PPM} no multileaved comparison method has been able to infer correct preferences without degrading the search experience of the user. In contrast, we prove that \ac{PPM} has both \emph{considerateness} and \emph{fidelity}, thus it is guaranteed to correctly identify a Pareto dominating ranker without altering the search experience considerably. Furthermore, our experimental results spanning ten datasets show that \ac{PPM} is more sensitive than existing methods, meaning that it can reach a lower level of error than any previous method. Moreover, our experiments show that the most significant improvements are obtained on the more complex datasets, i.e., larger datasets with more grades of relevance. Additionally, similar improvements are observed under different levels of noise and numbers of rankers in the comparison, indicating that \ac{PPM} is robust to interaction noise and scalable to large comparisons. As an extra benefit, the computational complexity of \ac{PPM} is polynomial and, unlike previous methods, does not depend on sampling or approximations.

The theoretical framework that we have introduced allows future research into multileaved comparison methods to guarantee improvements that generalize better than empirical results alone. In turn, properties like \emph{considerateness} can further stimulate the adoption of multileaved comparison methods in production environments; future work with real-world users may yield further insights into the effectiveness of the multileaving paradigm.
Rich interaction data enables the introduction of multileaved comparison methods that consider more than just clicks, as has been done for interleaving methods \cite{kharitonov2015generalized}. These methods could be extended to consider other signals such as \emph{dwell-time} or \emph{the order of clicks in an impression}, etc.

Furthermore, the field of \ac{OLTR} has depended on online evaluation from its inception \cite{yue09:inter}. The introduction of multileaving and following novel multileaved comparison methods brought substantial improvements to both fields \cite{schuth2016mgd, oosterhuis2016probabilistic}. Similarly, \ac{PPM} and any future extensions are likely to  benefit the \ac{OLTR} field too.

Finally, while the theoretical and empirical improvements of \ac{PPM} are convincing, future work should investigate whether the sensitivity can be made even stronger. For instance, it is possible to have clicks from which no preferences between rankers can be inferred. Can we devise a method that avoids such situations as much as possible without introducing any form of bias, thus increasing the sensitivity even further while maintaining theoretical guarantees?


\begin{spacing}{1}
\medskip\noindent\small
\textbf{Acknowledgments.}
This research was supported by
Ahold Delhaize,
Amsterdam Data Science,
the Bloomberg Research Grant program,
the Criteo Faculty Research Award program,
the Dutch national program COMMIT,
Elsevier,
the European Community's Seventh Framework Programme (FP7/2007-2013) under
grant agreement nr 312827 (VOX-Pol),
the Microsoft Research Ph.D.\ program,
the Netherlands Institute for Sound and Vision,
the Netherlands Organisation for Scientific Research (NWO)
under pro\-ject nrs
612.\-001.\-116, 
HOR-11-10, 
CI-14-25, 
652.\-002.\-001, 
612.\-001.\-551, 
652.\-001.\-003, 
and
Yandex.
All content represents the opinion of the authors, which is not necessarily shared or endorsed by their respective employers and/or sponsors.
\end{spacing}

\vspace*{-.5\baselineskip}

\balance

\bibliographystyle{abbrvnatnourl}
\bibliography{cikm2017-lambda-multileave}

\end{document}